\documentclass[runningheads]{llncs}
\usepackage{latexsym,amssymb,amsmath,latexsym}

\begin{document}

\title{Knowledge Representation in Agent's Logic with Uncertainty and Agent's Interaction}


\author{Maybin Muyeba$^a$ and Vladimir Rybakov$^{a,b-part-time}$}
\authorrunning{M.Muyeba and V.Rybakov}
\titlerunning{Knowledge Representation in Agent's Logic}

\institute{$^a$School of Computing, Mathematics and DT,
  Manchester Metropolitan University,
 John Dalton Building, Chester Street, Manchester M1 5GD, U.K. \\ and
 \\
  $^b$Institute of Mathematics,  Siberian Federal University,
 79 Svobodny Prospect, Krasnoyarsk, 660041, Russia
  \\
 \email{ e-mails: M.Muyeba@mmu.ac.uk and V.Rybakov@mmu.ac.uk}}

\toctitle{Lecture Notes in Computer Science}

\mainmatter

\maketitle



\newcommand{\pos}{\diamondsuit}
\newcommand{\cB}{\mathcal{B}}
\newcommand{\cK}{\mathcal{K}}
\newcommand{\cF}{\mathcal{F}}
\newcommand{\cm}{\mathcal{M}}
\newcommand{\cL}{\mathcal{L}}
\newcommand{\cC}{\mathcal{C}}
\newcommand{\cJ}{\mathcal{J}}
\newcommand{\bbZ}{\mathbb{Z}}
\newcommand{\vv}{\Vdash}
\newcommand{\nn}{\,\mathrm{Next}\,}  
\newcommand{\N}{\mathrm{N}\,}        
\newcommand{\pp}{\Diamond^+}
\newcommand{\uu}{\,\mathrm{Until}\,}
\newcommand{\und}{\,\&\,}
\newcommand{\Sk} {\mathrm{Sk}}
\newcommand{\Cl}{\mathop{Cl}\nolimits}
\newcommand{\Log}{\mathop{Log}}
\newcommand{\Ref}{\mathop{Ref}}
\newcommand{\Var}{\mathop{Var}}
\newcommand{\dom}{\mathop{dom}}
\newcommand{\Sub}{\mathop{Sub}}
\newcommand{\Fr} {\mathop{Fr}}
\newcommand{\logic}{\mathcal{T\!L}^{KnI,U}_{Dist}}
\newcommand{\llll}{\logic}
\newcommand{\td} {\mathrm{Today}}
\newcommand{\cz} {\mathcal{N}}
\newcommand{\mz} {\mathcal{N_C}}
\newcommand{\rf}[1]{{#1}_{\mathrm{rf}}}
\newcommand{\ppp}{\phi}






\begin{abstract}
This paper studies knowledge representation in multi-agent environment. We
 investigate technique for computation truth-values of statements based at
    new temporal, agent's-knowledge  logic $\logic$. A logical language, mathematical symbolic models and a temporal logic $\logic$ based at these models are suggested.
 We find an algorithm
which computes theorems of $\logic$ and satisfiability of statements, this implies that $\logic$ is
decidable (i.e. --  the satisfiability problem for $\logic$ is solvable).
Application areas are pointed and discussed.
\end{abstract}

\medskip


\ \ \ \ \ \ \ \ \ \ \ \ \ \ \ uncertainty, temporal logics,  decision algorithms

\medskip



\section{Introduction}

Knowledge may have difference origin and substance, but often it is assumed to be based at judgments of an
amount of agents. Actually we will  interpret knowledge as result/output of  interaction
between agents (it will be modeled by tools of a hybrid of multi-modal
agents' knowledge logic and temporal logic).
 Areas of applications for multi-agent systems (MAS) and knowledge based systems (KBS) are indeed utterly diverse, but
anyway, they are primarily focused to IT (Information Technologies) in various forms (cf. e.g. Badaracco  et al
\cite{DBLP:conf/kes/BadaraccoM11},
 K{\"o}nig et al \cite{h11a}, H{\aa}kansson \cite{h11},
H{\aa}kansson,  Hartung, \cite{h10} , H{\aa}kansson, \cite{h10b},
Burgin \cite{DBLP:conf/kes/BurginM11}, see also
\cite{h10c,h10d,h10e,DBLP:conf/kes/2011-2}).
Often logical instruments are useful, cf. eg.
  S.Cranefield \cite{sc4}
 considering  a logic for expression social expectations via conditional rules (individuals may be treated as agents with desired level of autonomy).

 Often  some variations of
modal and multi-modal logics are used for formalizing agent's reasoning.
Such logics
were, in particular, suggested in Balbiani et al \cite{bal1},
Vakarelov \cite{DBLP:conf/rsfdgrc/Vakarelov05}, Fagin et al
\cite{DBLP:journals/ai/FaginHV95,fag1}, Rybakov et al
\cite{vris,rit07,DBLP:conf/kes/BabenyshevR09}.
Representing probabilistic features of reasoning, often some elements of fuzzy logic
are efficiently implemented (cf. e.g. Ribaric et al \cite{flrib}). Working with implementation of various techniques in IT (for example in data mining) decision procedures and data elicitation again uses elements of logical reasoning (cf. e.g.  Muyeba et al \cite{mm1}).
In the paper Rybakov, Babenyshev \cite{vris} some multi-agent logic
modeling reasoning about distances in framework of temporal logic
was suggested, it proves decidability of this logic (and,
consequently, satisfiability problem).

This our current paper will study knowledge representation in logical terms.
We extend  results of \cite{vris}  to the case of a temporal multi-modal logic,
which also describes interaction of agents (knowledge by
interaction) and uncertainty. But we consider uncertainty  not via local knowledge, - as it was earlier in \cite{vris}, -- but now
via interaction of agents. We will use some extension of linear temporal  logic LTL
(cf. for LTL origin and applications
  Pnueli \cite{pnu},
 Manna and Pnueli~\cite{mp1},
  Barringer, Fisher, Gabbay and  Gough \cite{bf},
 Vardi~\cite{var3}).
  The mathematical theory
of temporal logics overall formed a highly technical branch in
the area of non-classical logic
 (cf.~van~Benthem~\cite{vb31,vb32},
  Gabbay and Hodkinson~\cite{gh1}, Hodkinson~\cite{hod1},
   de~Jongh et al.~\cite{ddj1}).

This our paper is devoted to computation of truth statements in multi-agent environment
within a temporal framework.
  We construct a new temporal logic $\logic$ defined in a semantic
 way via special Kripke models
  which describes frames where transition periods are filled
  with intermediate states.
   The
  satisfiability problem for this logic (or dually -- problem of
  decidability for $\logic$) is our prime aim.
 We find  an algorithm which computes theorems of logic $\logic$
(which implies that the logic is decidable, and the satisfiability problem
for it is also decidable).
 The general methodology of this paper
is borrowed from \cite{rit07} and
\cite{vris,DBLP:journals/logcom/Rybakov09}.

\section{Semantics for $\logic$, Modeling Runs of Time}

We, first, will introduce mathematical models for description agent's interaction
(and later on we will base a logic upon this semantics). We will use the following notation: 
 for
any set $A$ and a binary relation $R\subseteq A\times A$,

\begin{description}
    \item[$\bullet$]
$R^<$ be defined as follows:
 $a R^< b\iff aRb \und \neg(bRa)$;
      \medskip
    \item [$\bullet$]
  $R^2=R\circ R$, $R^{n+1}=R\circ R^n$ --- finite compositions of the relation $R$;
      \medskip
    \item[$\bullet$]
  $R^+=\bigcup_{n=1}^\infty R^n$ --- transitive closure of $R$;
      \medskip
    \item[$\bullet$]
  $R^*=\bigcup_{n=0}^\infty R^n$ --- reflexive and transitive
  closure of $R$.
    \end{description}

  A Kripke (multi-relation) frame
  $\langle C, R_1,\dots, R_m, R\rangle$, is a set $C$ with binary relations
  $R_1,\dots, R_m, R$.
  In the sequel, a \emph{multi-agent cluster} is a Kripke frame
  $\langle C,R_1,\dots, R_m, R\rangle$,
  where
  1) $R=C\times C$ is the universal relation on a set $C$;
  2) $R_1,\dots, R_m$ are equivalence relations on $C$.
    From this point on, we will call multi-agent clusters simply \emph{clusters},
  since we will not consider any other type of them.
  The class of all clusters we denote by $\Cl$.
    Given a cluster $C\in\Cl$, we denote
  $R_{1,C},\dots,R_{m,C},R_C$ the respective relations.
  A \emph{chain} is a frame
  $\langle\bigcup_{i=1}^n C_i,R_1,\dots, R_m, R\rangle$,
  where $C_1,\dots,C_n\in\Cl$ is a finite sequence
  of clusters, each $R_j$ is the union of individual
  $R_{j,C_i}$'s,
  and
  $  a R y \iff \exists i,j
  (i\leq j \und a\in C_i \und b\in C_{j}).
  $

    Let $\cC=C(0),C(1),C(2),\dots$ be
  a countable sequence of clusters.
    The basic semantic objects upon which we define our logic are the
Kripke models based on the following frames:

 \[
\mz :=
\left\langle
 \bigcup_{i\in N} C(i)\cup
\bigcup_{i\in N} \left\lfloor C(i),C(i+1)\right\rfloor,
R_1,\dots, R_m,
R, \nn
\right\rangle
\]
where
\begin{enumerate}
\item
  for each $i\in{N}$,
  $\left\lfloor C(i),C(i+1)\right\rfloor$
  is a collection (may be infinite) of chains 
  $\langle C_1,\dots,C_n\rangle$;
\item
each $R_j$, $j=1,\dots,m$ is the union
of the respective $R_{j,C}$, i.e.,
$$
R_j=
\bigcup_{i\in {N}} R_{j,C(i)}\cup
\bigcup_{i\in {N}}
\{ (R_{j,C}\mid
C\in\left\lfloor C(i),C(i+1)\right\rfloor\}
$$
\item
  $R=Q^+$, where
$$
\begin{array}{l}
Q=
\bigcup_{i\in {N}} R_{C(i)}\cup
\bigcup_{i\in {N}}
\{ R_{Chain}\mid
Chain\in\left\lfloor C(i),C(i+1)\right\rfloor\}\\[0.5em]
\hskip2em
\cup
\{\langle a,b\rangle \mid a\in C(i)\und b\in C_1\in \langle C_1,\dots,C_n\rangle\in
 \left\lfloor C(i),C(i+1)\right\rfloor\} \\[0.5em]
\hskip2em
\cup
\{
\langle a,b\rangle \mid
a\in C_n\in \langle C_1,\dots,C_n\rangle\in
 \left\lfloor C(i),C(i+1)\right\rfloor
\und
b\in C(i+1)
\}
\end{array}
$$
 \item
The relation $\nn$ is defined by
$$
\begin{array}{l}
a  \nn\ b  \iff \\[0.5em]
\hskip2em
(a\in C(i)\und b\in C(i+1))\\[0.5em]
\hskip2.5em
\vee
(a\in C\in Chain\in \lfloor C(i),C(i+1)\rfloor
\und b\in C(i+1)).
\end{array}
$$
\end{enumerate}

This semantics is similar to the one in
 \cite{vris}, but now we extended the language of the logic offered in
\cite{vris} to handle interactions the agents via uncertainty.

 \section{Syntax and  Language for $\logic$}

The logical language for our logic  $\logic$  contains usual temporal
operations $\nn$ (next) and $\uu$ (until), also we use new unary
logical operations $K_i$ for agent's knowledge, a special operator
$\td$, together with a countable set of operations for measuring
temporal distances $\{\pp_k\}_{k\in N}$.

Thus, the propositional language $\cL$ for $\logic$ includes the
following logical operations (logical connectives are given with
their arities as upper-right indices):
$$
\cL:=\langle
\vee^2,\wedge^2,\to^2,\neg^1,
\N^1,
\{K^1_i\}_{i=1}^m, \mbox{Unti$^1$},
\mbox{KnI$^1$}, \mbox{U}^1,
\{{\pp_k}^1\}_{k\in N},
\td^1,
\top^0,\bot^0\rangle
.$$

The alphabet of our logic uses propositional letters to denote not-identified statements: it  contains an enumerable set
 $\Var:=\{x_1,x_2,x_3,\dots\}$ of
 \emph{propositional variables}.
Formation rules for
 formulas over the propositional language $\cL$
are below:
$$
  \alpha
  ::=
  x_i\mathrel{|}
  \alpha_1\wedge\alpha_2\mathrel{|}
  \alpha_1\vee\alpha_2\mathrel{|}
  \alpha_1\to\alpha_2\mathrel{|}
  \neg\alpha\mathrel{|}
  K_i\alpha\mathrel{|}
  \N\alpha\mathrel{|}  $$

  $$  \alpha_1\uu\alpha_2\mathrel{|}
  \pp_k\alpha\mathrel{|} \mbox{KnI}\alpha \mathrel{|} \mbox{U}\alpha \mathrel{|}
  \top\mathrel{|}
  \bot 
 \ \ . $$

To describe this definition in less formal terms, this one means
(i) any propositional letter $x_i$ is a formula;
(ii) If $ \alpha_1$ and $\alpha_2$ are formulas than
$\alpha_1\wedge\alpha_2$, \
$\alpha_1\vee\alpha_2$, \
$\alpha_1\to\alpha_2$, \
    $ \neg\alpha$, \
    $K_i\alpha$, \
    $\N\alpha$, \
  $\alpha_1\uu\alpha_2$, \
    $\pp_k\alpha$, \
    $\mbox{KnI}\alpha$, \
    $ \mbox{U}\alpha$ \
    are again formulas.  $\top$ and $\bot$ (logical constants - true and false)
    are  formulas also.

\medskip

\section{Uncertain Statements, how we model}

 Now we would like to discuss the known approaches to handle logical uncertainty and to motivate our own approach. Maybe  a first approach to work with logical uncertainty was
 based at multi-valued logics (symbolic approach; studied since the 1920s as infinite-valued logics notably by Lukasiewicz and Tarski), and fuzzy logics (numerical approach; which, in more modern descent, may be referred to Lotfi A. Zadeh, mid 1960s).
Though, in such approaches, uncertainty is rather {\it directly} specified (so to say - enforced).
It is easy to confess that
 nobody can ever determine with an absolute certainty whether a proposition
 concerning a scientific doctrine or even statistic observations
   is true or false. Besides, whenever the truth of a statement is declared, it is always done by an individual,
   and it can never be considered to represent a general and objective belief (though social environment often inclines an individual to join to most popular
 viewpoint).
 
 \medskip
 
 {\bf Example.} Consider a network with an admin serving it and users for this network, -- as agents.
 Admit that these users and admin have an amount of assertions $\ppp$ about the state of this network
 (written in the language of suggested logic (coding eg. constancy, presence of specific errors, attempts to crack it, and so forth). How we cold determine that a statement $\ppp$ is uncertain?
 Consider the steps of inspections the network as a computation
 (indeed, the inspection may be undertaken by robots - software scripts - verifying some
 particular statements). Thus, agent's inspection is a computation, how then we may define uncertainty of 
 a statement $\ppp$?  There are several ways to approach it.
 For instance:
 
 \smallskip
 
\begin{itemize}
\item{(i)}
A statement $\ppp$ is uncertain if in  a future (after an interval of time in a computation) it will be a state when $\ppp$ is true and a state where $\ppp$ is false;

\vspace*{0.3cm}
\item{(ii)}
A statement $\ppp$ is uncertain if in current time cluster (e.g. -- in a tick of time while multi-thread computation, or in a web search in current time point, etc.) it is a state where $\ppp$ is true, and  it is  a state where $\ppp$ is false;

\vspace*{0.3cm}
\item{(iii)} To handle multi-agents' environment;
 a statement $\ppp$ is uncertain if some agent consider it to b true now, but an another one sees it is now false;

\vspace*{0.3cm}
\item{(iv)}
A statement $\ppp$ is uncertain if in the current time cluster (cf. (ii) for possible meaning) the following holds. Agents, passing information to each other (possible meaning: multi-thread computation and passing intermediate results via communication channels, communication of web admins via web pages available by admin logins and passwords (i.e access rules), human conversation by multiple phone calls, twitter etc) may achieve some state where $\ppp$ is true, and using similar, but another
procedure, they can find a state where $\ppp$ is false.
\end{itemize}

In current paper we will consider the case (iv) as most complicated and cute.
The other mentioned approaches also can be modeled in our technique, but we will consider (iv) not because it looks, so to say, most intricate. We think it reflects
 much better the essence of uncertainty (both from computational and philosophical
viewpoint) in agents' environment - via interaction of agents and a final conflict in opinions.

\section{Rules for computation of truth values for statements}

We turn now to description of rules
 for computation truth-values of formulas. For
any collection of propositional letters $Prop\subseteq\Var$ and any
frame $\mz$, a \emph{valuation} in  $\mz$ is a mapping, which
 assigns truth values to elements of $Prop$ in
$\mz$.   Thus, for any
 $p\in Prop$, $V(p)\subseteq \mz$.

  We will call any
$\langle \cz_C, V \rangle$ a (Kripke) model.
 For any such model $\cm$, the truth values can be extended from
propositions of $Prop$ to arbitrary formulas.
 For $a\in \cz_C$, we denote
$(\cm,a)\Vdash_V \ppp$ to say that the formula $\ppp$ is true at $a$ in
$\cm_C$ w.r.t. valuation $V$. Thus, $\forall p\in Prop$:
$(\cm,a)\Vdash_V p \iff a\in V(p)$.

Rules for computation truth-values for  boolean logical operations
are defined as usual, e.g.:
\( (\cm,a)\Vdash_V \ppp\wedge \psi \iff(\cm,a)\Vdash_V \ppp \wedge
(\cm,a)\Vdash_V \psi; \)
\( ((\cm,a)\Vdash_V \neg \ppp \ \iff \) \(\ not [(\cm,a)\Vdash_V \ppp]. \)

For other logical operations, suppose $a,b\in \cz_C$. Then

    \((\cm,a)\Vdash_V K_i \ppp
      \iff \forall b (aR_i b\implies (\cm,b)\Vdash_V \ppp)\);
\ \ \
    \((\cm,a)\Vdash_V \N \ppp
      \iff \forall b\,
    (a \nn b \implies (\cm,b)\Vdash_V \ppp)\);
\ \ \   \((\cm,a)\Vdash_V \phi\uu \psi
             \iff
    \exists b\, (a\,\mathrm{Next}^* b\und (\cm,b)\Vdash_V \psi
         \und\)
         \( \forall c (a\,\mathrm{Next}^* c \,\mathrm{Next}^+ b{\implies}
    (\cm,c)\Vdash_V \ppp)\);
      \ \ \
    \((\cm,a) \Vdash_V \pp_k\ppp \iff
    \exists b ( a (R^<)^k b\und (\cm,b)\Vdash_V \ppp)\);
\ \ \
    \((\cm,a) \Vdash_V \td\ppp \iff
    \forall b\in C(a)\, ((\cm,b)\Vdash_V \ppp)\);
\ \ \
        \((\cm,a)\Vdash_V \mbox{KnI}\ppp \ \iff
  \ \exists a_{i1}, a_{i2}, \dots , a_{ik} \in C(a)\)
   \( [
  a R_{i1} a_{i1} R_{i2} a_{i2} \dots R_{ik} a_{ik}]
   \& \ \&  (\cm,a_{ik})\Vdash_V
\ppp \).
An important step in our approach is definition of logical uncertainty:

\[(\cm,a)\Vdash_V \mbox{U} \phi  \  \iff  \mbox{KnI}\ppp \wedge \mbox{KnI} \neg \ppp.
\]

So, we assume that the logical truth of a statement $\ppp$ is uncertain if agents may know via
own interaction and passing knowledge one to other that $\ppp$ may be true and also $\ppp$ may be false.
Usage of the logical operations has the following not-formal meaning:
\begin{itemize}
    \item
\((\cm,a)\Vdash_V K_i \ppp\)  --- at the state $a$
 agent $i$ knows that $\phi$;
  \medskip
    \item
\((\cm,a)\Vdash_V \td\, \ppp\)  ---
  $\phi$ holds today (relatively to the time moment of $a$);
  \medskip
    \item
\((\cm,a)\Vdash_V \N \ppp\)  ---
  $\phi$ holds tomorrow (counting from $a$);
  \medskip
    \item
\((\cm,a)\Vdash_V \pp_k \phi\)  ---
  $\phi$ holds in $k$ steps from now (counting from  $a$).

  \item \vspace*{0.3cm}
$(\cm,a)\Vdash_V \mbox{KnI} \phi$ --- in the current state $a$, the statement
$\phi$ {\it may be
known by interaction between agents};
\item \vspace*{0.3cm}
$(\cm,a)\Vdash_V \mbox{U} \phi$ --- the statement $\phi$ is uncertain at the state $a$.
\end{itemize}

\begin{definition}
$$
\mbox{Logic $\llll$ is the set of all formulas
 which are valid in all frames $\mz$.}
$$
\end{definition}

\medskip

We say a formula $\varphi$ is a theorem of $\logic$ if $ \varphi \in \logic $; a formula $\varphi$ is satisfiable if
there is frame and a valuation in this frame such that $\varphi$ is true at some state of this frame.
Theorems of $\logic$ are valid statements, -- i.e. those which are always true.

          \medskip

{\bf Possible applications of the semantics and suggested logic.}
This semantics may be applied for modeling various reasoning (decision making) concerning multi-agent environment.
It could be multi-threads  computations with intermediate channels for exchanging current results of computation.
Some environment close to human reasoning involving many individuals, -- as web and phone conferences, --
 good matches to accepted formalism as well. Another application areas could be web search for information via multiple web pages by
 many
individuals with shared or distributed access rules.
Any area, where communication and interaction of agents assumes passing intermediate information, can be efficiently modeled in our framework.

\section{Satisfiability and Decidability, Computing Algorithms}

Now we turn  to computational problems for our suggested logical system.
How to compute that a given statement is true, how to see that it is satisfiable, how to decide it?
We will use techniques to handle
inference rules from \cite{rint,rit071,vr11,vrlc}, since it works very well for our aims. To recall necessary definitions,
an inference rule is
 a relation
 \[ {r}:= \frac{\varphi_1( x_1, \dots ,
x_n), \dots , \varphi_l( x_1, \dots , x_n)}{\psi(x_1, \dots , x_n)},
\]
where $\varphi_1(x_1, \dots , x_n), \dots , \varphi_l( x_1, \dots ,
x_n)$ and $\psi(x_1, \dots , x_n)$ are
 formulas constructed out of
letters $x_1, \dots , x_n$.
The letters $x_1, \dots , x_n$ are the
variables of ${r}$, we use the notation $x_i \in Var(r)$.

Informal meaning of this rule is: $\varphi_1( x_1, \dots ,
x_n), \dots , \varphi_l( x_1, \dots , x_n)$ are premisses (assumptions) and $\psi(x_1, \dots , x_n)$ is the conclusion of $r$:
$r$ says that the conclusion follows from assumptions.

A rule
  $r$ is said to be \emph{valid} in
a Kripke model $\langle {\cz_C}, V \rangle$
(we will use notation  ${\cz_C}\Vdash_V {r}$) if
 \[
\forall a \ (({\cz_C,a}) \Vdash_V \bigwedge_{1\leq i \leq
l}\varphi_i)
\implies \forall a \ (({\cz_C},a) \Vdash_V \psi).
\]
Otherwise we say ${r}$ is \emph{refuted} in $\cz_C$ (or
\emph{refuted in $\mz$ by $V$}), and write ${\mz}\not\Vdash_V {r}$. A
rule ${r}$ is \emph{valid} in a frame ${\cz_C}$ (notationally,
${\cz_C }\Vdash {r}$) if, for any valuation $V$, ${\cz_C}\Vdash_V {r}$.
Since our language $\cL$ includes conjunction we can consider only
rules with one-formula premise.

 Being given with an arbitrary formula $\ppp$, we can convert $\ppp$ into the
  rule $x \to x / \ppp$
  and employ a technique of reduced normal forms for inference
rules as follows. The following statement immediately follows from definitions.

\begin{lemma} \label{p1}
A formula $\ppp$ is a theorem of $\llll$
iff the rule $({x \to x / \ppp})$ is valid in any frame $\mz$.
\end{lemma}

So, instead of theorems we may consider valid inference rules.

 A rule $\rf{r}$  is said to be in
\emph{reduced normal form} if
$\rf{r}= \bigvee_{1\leq j \leq s}\theta_j / x_1$,
where each $\theta_j$ has the form:
$$
\begin{array}{l}
\displaystyle
\theta_j =
 \bigwedge_{i=1}^n
  x_i^{t(j,i,0)}
  \wedge
 \bigwedge_{i=1}^n
  (\nn x_i)^{t(j,i,1)}
  \wedge
 \bigwedge_{i=1;l=1}^{n;m}
   (K_l x_i)^{t(j,l,i,1)}
  \\[0.5em]
  \hskip4em
\displaystyle
  \wedge
 \bigwedge_{i=1;l=0}^{n;k}
  (\pp_l x_i)^{t(j,i,l,2)}
  \wedge
 \bigwedge_{i,l=1}^n
  (x_i \uu x_l)^{t(j,i,l,3)}
  \\[0.5em]
  \hskip4em
\displaystyle
  \wedge ~~\mbox{KnI} x_i^{t(j,i,2)} \wedge ~\mbox{U} x_i^{t(j,i,3)}
\end{array}
$$
for some values $t(j,i,z), t(j,i,k,z)\in \{0,1\}$
and where, for every formula $\alpha$ above,
 $\alpha^0:=\neg\alpha$, $\alpha^1:= \alpha$.

For a rule $\rf{r}$ in the reduced normal form,
$\rf{r}$
is said to be a \emph{normal reduced form for a rule $r$} iff,
for any frame $\mz$,
\[\mz \Vdash r
\iff \mz \Vdash \rf{r}.\]

  Based at  the technique similar to one described in
  \cite[Section~3.1]{ry97}, we can transform
  every inference rule in the language $\cL$
to a definably equivalent rule in the reduced normal form.

\begin{lemma}
\label{mt3}
 Every rule $r=\alpha/\beta$ can be transformed in exponential time
 to a definably equivalent
 rule $\rf{r}$ in the reduced normal form.
\end{lemma}

We use this lemma for algorithms to solve satisfiability and decidability of $\llll$.
Notice that a formula $\ppp$ is satisfiable iff $\ppp$ is not a theorem of $\logic$.

So, decidability implicates solution for satisfiability problem.
The decidability of $\llll$ will follow (by Lemma~\ref{p1}) if we
 find an algorithm recognizing rules in the
reduced normal form which are valid in all frames $\cz_C$.
For our
approach to complete our scheme, we need  one more construction of
special Kripke frames; in a sense, they are looking similar to
frames $\cz_C$, but have a bit another structure.
The structure of these frames is more complicated in comparison with used $\cz_C$ and we will omit their detail description
due to size of paper limitation.
 We will denote such frames by
$\cz^{\,\sharp}_C$. To complete our approach we will  apply proof scheme from Rybakov \& Babenyshev
\cite{vris} specified more in  \cite{DBLP:journals/logcom/Rybakov09}.

\begin{lemma} \label{oo11} {\sl A rule $\bf r_{nf}$ in reduced
normal form is refuted in a frame $\cz_C$ if and only if $\bf r_{nf}$
can be refuted in a frame of the same sort but with clusters of size square
exponential from $\bf r_{nf}$.}
\end{lemma}

Using this lemma and structure of frames $\cz^{\,\sharp}_C$,
following closely to proof from \cite{vris},
we derive

\begin{lemma} \label{oo1}
 A rule $\rf{r}$ in  the reduced
normal form is refuted in a frame $\cz_C$ iff $\rf{r}$ can be
refuted in some frame $\cz^{\,\sharp}_C$ by a valuation $V$ of
special kind, where the size of the frame
$\cz^{\,\sharp}_C$ is effectively computable from $\rf{r}$.
\end{lemma}

Now, using
 Theorem \ref{mt3}, Lemma \ref{p1} and Lemma \ref{oo1} we obtain

\begin{theorem}
\label{bn1} The logic $\llll$ is decidable. The algorithm for
checking a formula $\ppp$ to be a theorem of logic $\llll$ consists
of verification for validity rules in the reduced normal form at
frames $\cz^{\,\sharp}_C$ of size s effectively computable from the size of the
formula $\ppp$.
\end{theorem}

As we noticed above, this theorem  gives an algorithm which checks satisfiability in $\logic$.
The algorithm is based at construction of the frame $\cz^{\,\sharp}_C$ in Lemma \ref{oo1}.

\section{Conclusions and Future Work}

Our
 technique might be useful in applications to many areas.
For example, it may efficiently work in study and modeling
 reasoning (decision making) concerning various multi-agent environments.
A good example is
 multi-threads  computations (where any thread is an agent) with intermediate channels for exchanging by current results of computations.
Reasoning, discussions at web and phone conferences (actually any remote conversation (discussion)) may be
in reasonable depth formalized in suggested technique.
Web (database) search for information via multiple web pages (several/many databases) by
 sets of
individuals or web robots (as agents) with shared or distributed access rules
 very well suits again for modeling in our suggested framework.

There are many prospective avenues to continue this research.
First of all the suggested technique has not too good computational efficiency as it uses
computation of truth values for rules in models -- which is computationally very costly.
Thus,  improvements of computational efficiency would be very desirable.
Next interesting problem is to transfer the suggested approach to non-linear temporal logics: the cases when running of time (computational threads)
is not linear.

\bibliographystyle{plain}
\bibliography{mal}

\end{document}